# Correction of broadband terahertz electro-optic sampling with GaSe crystals

Kotaro Ogawa,[1,*] Natsuki Kanda,[1,2] Yuta Murotani,[1] Shunsuke Tanaka,[1,3] Jun Yoshinobu,[1] Ryusuke Matsunaga[1]

[1] *The Institute for Solid State Physics, The University of Tokyo, 5-1-5 Kashiwanoha, Kashiwa, Chiba 277-8581, Japan.*
[2] *RIKEN Center for Advanced Photonics, RIKEN, 2-1 Hirosawa, Wako, Saitama 351-0198, Japan.*
[3] *National Institute of Advanced Industrial Science and Technology, 1-1-1 Umezono, Tsukuba, Ibaraki 305-8563, Japan.*
*\*ogawa_k@issp.u-tokyo.ac.jp*

**Abstract: Gallium selenide (GaSe) is an efficient nonlinear crystal for electro-optic (EO) sampling in the multi-terahertz (THz) frequency range. However, the lattice resonance at several THz frequencies hampers broadband EO sampling, resulting in distorted pulse waveforms. In this work, we experimentally evaluated the frequency-dependent response function in EO sampling, considering the effects of phonons, phase mismatch, and gate pulse waveforms. The phonon effect is described using an effective Faust-Henry coefficient, which was determined to be -0.21 ± 0.02. The corrected field amplitude of multi-THz pulses aligns with additional measurements of average power and a beam diameter. The successful compensation of the frequency characteristics in GaSe will contribute to a more accurate evaluation of multi-THz transients.**

Femtosecond (~100 fs) pulsed lasers enable THz pulse generation and time-domain detection using electro-optic (EO) crystals, which has spurred intensive research into observing and controlling the low-energy properties of materials [1–4]. By using ultrashort pulses close to 10 fs, conventional THz time-domain spectroscopy (THz-TDS) can be extended to higher frequency ranges, up to several tens of THz [5,6]. In this multi-THz range, an intense peak electric field as large as 100 MV/cm is achievable in a tabletop setup [7]. In addition, broadband spectra of multi-THz pulses generated by EO crystals [8,9], two-color laser-induced air plasma (LIAP) [10–12], and spintronic emitters [13] serve as ultrafast driving fields for nonthermal control of matter, as well as a novel probe covering the entire range of infrared responses [14–17]. Furthermore, artificial shaping methods of multi-THz waveforms, such as slicing with a plasma shutter [18] and counterrotating bicircular light [19], have enabled the realization of unique transients in the time domain, which will advance ultrafast electronics and photonics with designed light fields [20,21].

For these purposes, an accurate sampling method for field transients spanning the THz to multi-THz bands is required. One of the most commonly employed methods is EO sampling, which is simple, highly sensitive, and capable of polarization-sensitive measurements [22]. However, a broadband multi-THz pulse waveform detected by EO sampling is significantly distorted due to a detection efficiency "gap" arising from optical phonon resonances in EO crystals [23,24]. Heterodyne air-biased coherent detection can achieve distortion-free broadband sampling [10], but requires intense gate pulses and a high bias voltage. Recently, gapless broadband THz sampling beyond 15 THz in bias-free heterodyne detection was demonstrated using metal surfaces as a local oscillator without suffering from the phonon resonance [25]. Zeeman torque in a ferromagnetic metal [26] and spin accumulation at a ferromagnetic-metal/heavy-metal interface [27] were also employed for gapless sampling. However, these responses in metals are still less efficient than EO sampling. For practical use of gapless sampling for broadband multi-THz pulses, proper correction of the phonon effect on EO sampling is necessary.

In conventional THz-TDS, the principle of EO sampling is described by the Pockels effect, where a static electric field modifies the refractive index of EO crystals. The balanced-detection signal is given by $S(t) = \Omega_0 d n_0^3 r_{\text{eff}}^e I_0 E_{\text{THz}}(t)/c$ [28], where $t$ is the delay time, $\Omega_0$ the gate pulse's central frequency, $d$ the crystal thickness, $n_0$ the refractive index at $\Omega_0$, $r_{\text{eff}}^e$ a nonlinear optical coefficient of the EO crystal, $I_0$ the gate pulse intensity, $E_{\text{THz}}(t)$ the incident THz electric field, and $c$ the speed of light. This expression is based on several assumptions: (i) the gate pulse duration is short enough to treat the THz field as static, (ii) the refractive index of the EO crystal is independent of frequency, (iii) phase mismatch during propagation inside the crystal is negligible, and (iv) the optical frequency of the gate pulse is represented by a single value $\Omega_0$. In the multi-THz regime, however, the optical phonons, typically appearing at several THz, induce a considerable frequency dependence of the optical coefficient, thus altering the phase-matching condition. Furthermore, ultrashort gate pulses have a broad bandwidth, making their spectral information relevant to the EO sampling signal. Previous studies have demonstrated the correction of the phonon effect for GaP and ZnTe [23,24], typical EO crystals in the sub- to a few-THz regions. To model the influence of phonons, the Faust-Henry (FH) coefficient $C_{\text{FH}}$, which represents the ratio between lattice and electron contributions to the EO effect [29], has been included in the analysis. To extend this method to higher frequencies, the FH coefficient for GaSe, a representative EO crystal in the multi-THz band (Fig. 1(a)), is indispensable [30]. A conventional method to determine the FH coefficient is Raman spectroscopy [29,31], which has suggested $C_{\text{FH}} = -0.37$ for GaSe [32,33]. However, the simple relationship between the EO effect and spontaneous Raman scattering may be affected by extrinsic factors, such as surface quality and propagation of light, possibly leading to a discrepancy between the predicted and measured values. For ZnTe, a similar discrepancy was noted between theoretical prediction ($C_{\text{FH}} = -0.32$) [34] and the value measured by THz-TDS ($C_{\text{FH}} = -0.07$) [24]. To fully compensate for the phonon effect in the EO sampling, an *effective* FH coefficient $C_{\text{FH}}^{\text{eff}}$ should be directly determined in THz-TDS by using a targeted EO crystal to detect a broadband, gapless THz pulse and by carefully analyzing the effect of phonons in the amplitude and phase spectra.

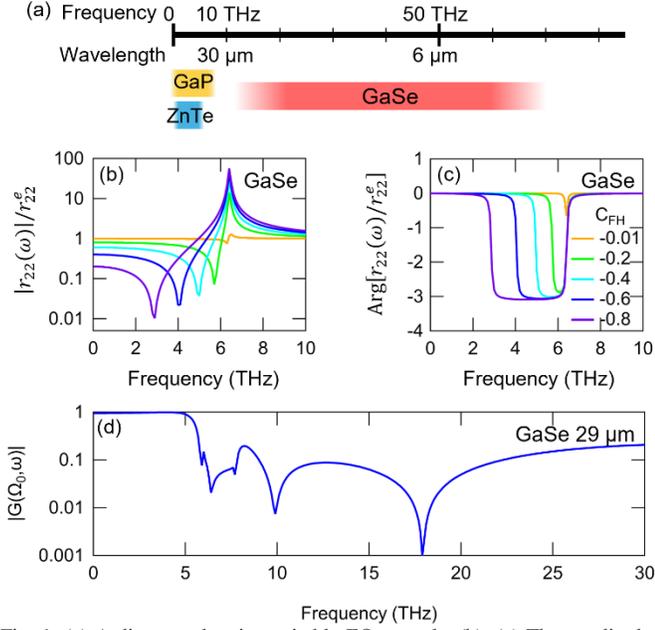

Fig. 1. (a) A diagram showing suitable EO crystals. (b), (c) The amplitude and phase spectra of $r_{22}(\omega)/r_{22}^e$ for GaSe with different $C_{\text{FH}}^{\text{eff}}$ values from $-0.01$ to $-0.8$. (d) The amplitude spectrum of the phase-mismatch function $G(\Omega_0, \omega)$ calculated at $\Omega = \Omega_0$ and $d = 29$ μm for GaSe.

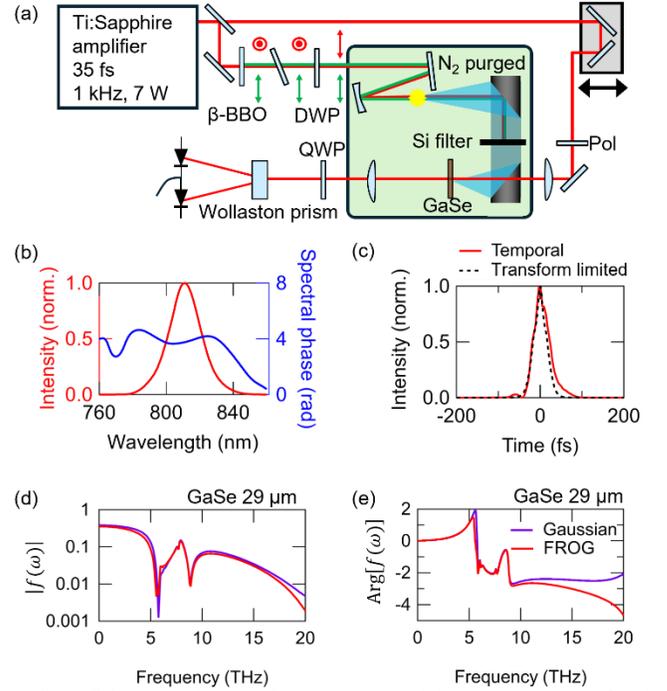

Fig. 2. (a) Schematic of THz pulse generation and detection using LIAP and EO sampling. (b) Retrieved intensity and phase spectra of the gate pulse from FROG analysis. (c) Retrieved temporal intensity profile and its transform-limited pulse. (d), (e) Amplitude and phase spectra of the response function for the actual gate pulses with complex spectra (FROG) and flat-phased ideal Gaussian gate pulses with $C_{\text{FH}}^{\text{eff}} = -0.23$.

In this letter, we demonstrate the compensation of the phonon effect on the EO sampling response function of GaSe by evaluating the effective FH coefficient using broadband pulses emitted from LIAP. The corrected field amplitude obtained using this method shows reasonable agreement with complementary measurements from a power meter and an IR camera. This result paves the way for the accurate evaluation of broadband multi-THz pulses, which could potentially lead to distortion-free EO sampling spanning THz to multi-THz bands.

The relation between the balanced signal in the EO sampling using GaSe and a THz field $E_{\text{THz}}$ can be more accurately expressed as

$$S(t) = \frac{n_0^3 \Omega_0 d r_{22}^e I_0}{c} \int_{-\infty}^{\infty} f(\omega) E_{\text{THz}}(\omega) e^{-i\omega t} d\omega, \quad (1)$$

where $\omega$ is THz frequency, and $\Omega_0$ the central frequency of the gate pulse. $r_{22}(\omega) = r_{22}^e \times \left(1 + C_{\text{FH}}^{\text{eff}}\{1 - [(\hbar\omega)^2 + i\hbar\omega\gamma]/(\hbar\omega_{\text{TO}})^2\}^{-1}\right)$ is an EO coefficient around the phonon resonance [29]. We used $\hbar\omega_{\text{TO}} = 213.5$ cm$^{-1}$ ($= 6.4$ THz) and $\gamma = 3$ cm$^{-1}$ for GaSe [35]. $f(\omega)$ is the response function that can be obtained by solving the wave equations for EO sampling [36] as

$$f(\omega) = \frac{2 r_{22}(\omega)}{[1 + \tilde{n}(\omega)] r_{22}^e I_0} \int_{-\infty}^{\infty} T(\Omega) G(\Omega, \omega) I(\Omega, \omega) d\Omega. \quad (2)$$

$\tilde{n} = n + i\kappa$ a complex refractive index for the gate pulse. $T(\Omega) = n(\Omega_0)\Omega/n(\Omega)\Omega_0 \times \exp[-2\kappa(\Omega)\Omega d/c]$ accounts for propagation of the gate pulse in the EO crystal. $G(\Omega, \omega) = (\exp[i\Delta k(\Omega, \omega)d] - 1)/i\Delta k(\Omega, \omega)d$ represents the effect of phase mismatch $\Delta k(\Omega, \omega)$. $I(\Omega, \omega) = E_{\text{gate}}^*(\Omega) E_{\text{gate}}(\Omega - \omega)$ is an autocorrelation of the electric field of the gate pulse. Figures 1(b) and 1(c) show the amplitude and phase spectra of $r_{22}(\omega)/r_{22}^e$ when $C_{\text{FH}}^{\text{eff}}$ varies from $-0.01$ to $-0.8$, indicating that the dip in the amplitude and the phase jump are sensitive to changes in $C_{\text{FH}}^{\text{eff}}$. Figure 1(d) shows the phase mismatch function $G(\Omega_0, \omega)$ for the case of $\Omega = \Omega_0$ and $d = 29$ μm, which also exhibits a strong phonon influence around 6 THz. The large dips at 10 and 18 THz originate from the phase mismatch and are sensitive to $d$, which is discussed in a subsequent paragraph.

Owing to the sharp dependence of $r_{22}(\omega)$ on $C_{\text{FH}}^{\text{eff}}$, the response function $f(\omega)$ also exhibits a sensitive change in its frequency response (Supplement 1). This behavior allows us to evaluate $C_{\text{FH}}^{\text{eff}}$ by (i) preparing a gapless broadband THz pulse source with a smooth spectrum around the phonon resonance at 6.4 THz, (ii) detecting the pulse via EO sampling with GaSe, and (iii) determining the value of $C_{\text{FH}}^{\text{eff}}$ that smoothly corrects the spectrum. We utilized broadband THz pulses generated by LIAP, which serve as a gapless light source owing to the absence of infrared-active vibrational modes in the generation process. The optical system is illustrated in Fig. 2(a). A Ti:Sapphire regenerative amplifier system (Solstice Ace, MKS Instruments, Inc.) with a pulse energy of 7 mJ, a central wavelength of 811 nm, a repetition rate of 1 kHz, and a pulse duration of 35 fs was used as the light source. The output was separated by a beam splitter with an intensity ratio of 9:1 for THz pulse generation and the gate pulse. The former was directed into a

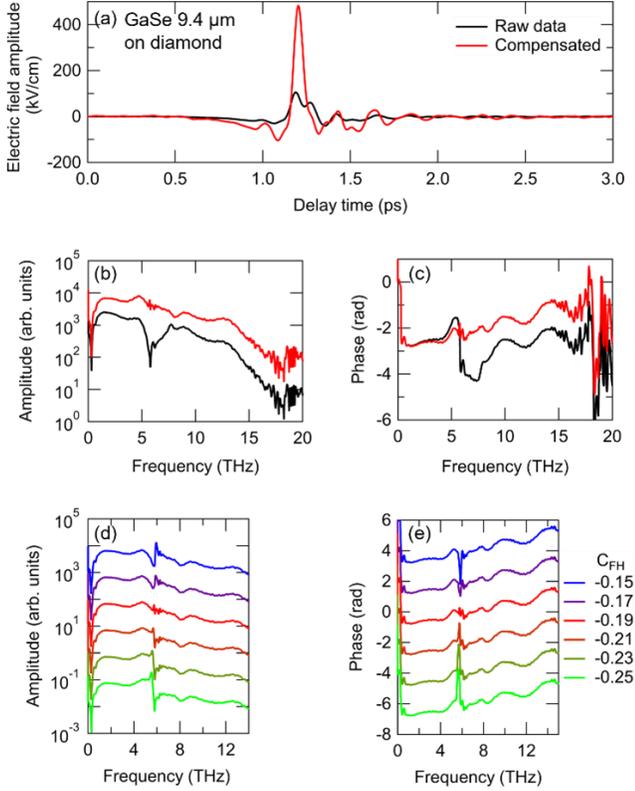

Fig. 3. The temporal waveforms (a), amplitude spectra (b), and phase spectra (c) of the THz pulse generated by LIAP. The black curves were raw data obtained by a 9.4 µm-thick GaSe on a diamond substrate, and the red curves were corrected by the response function with $C_{\text{FH}}^{\text{eff}} = -0.19$. (d), (e) The corrected amplitude and phase spectra for various values of $C_{\text{FH}}^{\text{eff}}$ with vertical offsets.

β-BaB$_2$O$_4$ (BBO) crystal to generate second harmonic light with type-I phase matching. The relative phase between the two colors was adjusted by rotating a thin calcite plate. A dual waveplate (DWP) was used to align the polarizations of the two colors collinearly. The two-color light was focused in an N$_2$-purged box to induce filamentation. A high-resistivity 1 mm-thick Si wafer was inserted as a filter, blocking NIR while transmitting THz. We confirmed that the transmittance of the Si filter was nearly constant below 12 THz (Supplement 2). The THz pulses emitted by LIAP were passed through the Si filter and detected by EO sampling in a GaSe crystal, using gate pulses and a balanced detection scheme.

The analysis in Eq. (2) requires the complex electric field spectrum of the gate pulse, including its phase information. Therefore, the complex electric field spectrum was measured using the second-harmonic generation-based frequency-resolved optical gating (SHG-FROG) method. The corresponding FROG traces are shown in Supplement 3. Figure 2(b) presents the retrieved intensity and phase spectra. Figure 2(c) shows the temporal intensity profile with a duration of 42.3 fs, which is slightly longer than that of the transform-limited pulse with a 35.2-fs duration, indicating an available bandwidth up to ~23 THz. Figures 2(d) and 2(e) display the amplitude and phase spectra of the calculated response function $f(\omega)$ for $d = 29$ µm and $C_{\text{FH}}^{\text{eff}} = -0.23$. To examine the effect of dispersion in the gate pulses, we compare the response functions of the actual gate pulses evaluated by FROG and the transform-limited Gaussian pulses with the same duration. The frequency dependence of detection sensitivity is affected by the distorted shape of the gate pulse waveform and intensity spectrum. While there is considerable deviation above 20 THz, the results agree well below 20 THz. The findings indicate that the phase information of the gate pulses plays only a minor role in the region of interest.

The black curve in Fig. 3(a) shows the raw data of the EO sampling waveform for the THz wave generated by LIAP. The EO crystal was a 9.4 µm-thick GaSe cleaved in the *ab*-plane. To suppress backside reflection, the GaSe crystal was mounted on a 500 µm-thick diamond substrate with a refractive index close to that of GaSe in the THz range [37]. The field strength for the raw data (black) was estimated as $E_{\text{THz}} = (\lambda_0 / 2\pi d n_0^3 r_{22}) \times \Delta I / I_0$. The amplitude and phase spectra of the pulse are shown as black curves in Figs. 3(b) and 3(c), where a sharp dip in the amplitude and an abrupt phase shift are observed in the range 6.4–7.6 THz. These features are attributed to the phonon resonance in GaSe. The amplitude signal begins to drop above 14 THz, which is due to phonon absorption in the Si filter as well as the time resolution of the 42.3 fs-long gate pulse. The actual spectrum of the pulses emitted from LIAP extends into the mid-infrared [12].

To evaluate an effective FH coefficient $C_{\text{FH}}^{\text{eff}}$, the corrected spectrum $E_{\text{THz}}(\omega) = S_{\text{THz}}(\omega)/f(\omega)$ was calculated using various values of $C_{\text{FH}}^{\text{eff}}$. The results of the amplitude and phase spectra are shown in Figs. 3(d) and 3(e), where $C_{\text{FH}}^{\text{eff}}$ was varied between $-0.25$ and $-0.15$. The distortions in the amplitude and phase spectra near the phonon resonance were smoothly compensated when $C_{\text{FH}}^{\text{eff}} = -0.19$ as shown by the red curves in Figs. 3(d) and 3(e). The corresponding corrected amplitude and phase spectra were compared with the raw data in Figs. 3(b) and 3(c), shown as the red curves without vertical offset. The increase in amplitude on the low-frequency side is mainly attributed to the correction of the Fresnel transmittance $2/[1 + \tilde{n}(\omega)]$. The oscillation in the 10–15 THz range can be attributed to interference from residual backside reflections at the GaSe-substrate interface. From the inverse Fourier transform analysis, the corrected electric field waveform was obtained, as shown by the red curve in Fig. 3(a). The peak electric field was evaluated as $4.8 \times 10^2$ kV/cm, five times larger than that of the raw data (black).

These results indicate that the response functions with $C_{\text{FH}}^{\text{eff}} = -0.19$ can smoothly compensate for the frequency characteristics of EO sampling signals for the 9.4 µm-thick GaSe on the diamond substrate. Although the FH coefficient itself is an intrinsic parameter of EO crystals, the value of $C_{\text{FH}}^{\text{eff}}$ empirically determined in this way could include extrinsic effects such as lattice distortion and surface quality of the crystal. We examined several different spots for focusing the gate pulses on the same GaSe crystal and confirmed that the value of $C_{\text{FH}}^{\text{eff}}$ for optimal correction did not change. To verify the sample dependence, we performed a similar procedure for another free-standing bulk GaSe crystal with a thickness of 29 μm. The black curves in Figs. 4(a) and 4(b) show the raw temporal waveform and amplitude spectrum, respectively, for the broadband THz pulses. The time-domain waveform exhibits a backside reflection after the main peak, which generates interference fringes in the amplitude spectrum. In the amplitude spectrum, two large dips appear at approximately 6 and 9 THz; the former is due to the phonon resonance, and the latter originates from phase mismatch, as shown in Fig. 1(d). This behavior allowed for precise determination of $C_{\text{FH}}^{\text{eff}}$ such that the two dips were most smoothly compensated (Supplement 4). The results of compensation using $C_{\text{FH}}^{\text{eff}} = -0.23$ and $d = 29.0$ μm are shown by the red lines in Fig. 4. The peak value of the temporal waveform was evaluated to be $4.7 \times 10^2$ kV/cm, successfully reproducing the result shown in Fig. 3(a). $C_{\text{FH}}^{\text{eff}} = -0.23$ for the free-standing 29.0 μm-thick GaSe differs slightly from $C_{\text{FH}}^{\text{eff}} = -0.19$ for the 9.4 μm-thick GaSe on the substrate. The deviation may be attributed to strain induced by the substrate.

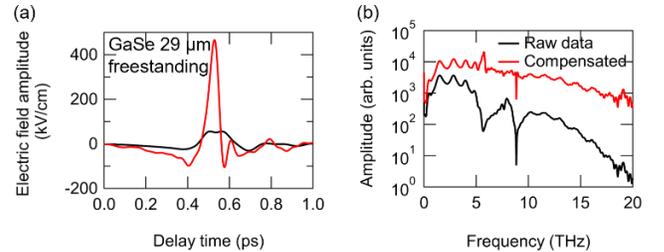

Fig. 4. (a) Temporal waveform and (b) amplitude spectrum of the THz electric field. The black curves were measured using the free-standing GaSe crystal, and the red curves were compensated using the response function with $C_{\text{FH}}^{\text{eff}} = -0.23$ and $d = 29.0$ µm.

To confirm the validity of this compensation, the peak electric field was also evaluated based on (i) pulse energy, (ii) pulse width, and (iii) beam diameter of the generated THz pulse. The pulse energy at the focal point was measured with a power meter (3A-P-THz, Ophir) and found to be 2.0 μJ in the 2–10 THz range. The pulse width was obtained from the raw temporal waveform (black curve in Fig. 3(a)) and calculated as the full width at half maximum of the Hilbert-transformed envelope, yielding 76 fs. The beam diameter was estimated using an MIR camera (RIGI-M2, Swiss Terahertz) with sensitivity in the 16.7–75 THz range. Since the THz pulse peaks at approximately 1 THz, we approximated the beam waist at this frequency by assuming a frequency-dependent beam size, obtaining a diameter of 4 mm (Supplement 5). Using these values, the peak electric field was estimated as $5 \times 10^2$ kV/cm, which is in reasonable agreement with the corrected EO sampling in Figs. 3(a) and 4(a).

In conclusion, we evaluated the effective FH coefficient of GaSe crystals to correct the frequency response of the EO signals. The electric field waveforms were successfully reconstructed, and the peak electric field obtained from the correction is in reasonable agreement with the values measured by a power meter and an IR camera. This result paves the way for distortion-free characterization of ultrabroadband infrared pulses based on the efficient EO response spanning the THz to multi-THz range.

**Funding.** JST PRESTO (Grant No. JPMJPR2006); JST FOREST (Grant No. JPMJFR2240); JST CREST (Grant No. JPMJCR20R4); JSPS KAKENHI (Grants No. JP24K00550 and No. JP24K16988); MEXT Q-LEAP (Grant No. JPMXS0118068681).

**Disclosures.** The authors declare no competing interests.

**Data Availability.** Data underlying the results presented in this paper are not publicly available but may be obtained from the authors upon reasonable request.

**Supplemental Document.** See Supplemental information for supporting content.